\RequirePackage{fix-cm}
\documentclass[twocolumn]{svjour3}          
\smartqed  
\usepackage{graphicx}
\usepackage{amsmath,amssymb}
\newcommand{\be}{\begin{equation}}
\newcommand{\ee}{\end{equation}}
\newcommand{\mk}{\langle k\rangle}
\newcommand{\bk}{\mathbf{k}}
\journalname{Biological Cybernetics}
\begin{document}

%
%

\title{The effects of within-neuron degree correlations
 in networks of spiking neurons 
}


\author{Carlo R. Laing        \and
        Christian Bl{\"a}sche 
}


\institute{Carlo R. Laing \at
School of Natural and Computational Sciences, 
Massey University, Private Bag 102-904 NSMC, Auckland, New Zealand. \\
phone: +64-9-414 0800 extn. 43512
fax: +64-9-4418136 \\
              \email{c.r.laing@massey.ac.nz}           
           \and
           Christian Bl{\"a}sche \at
             School of Natural and Computational Sciences, 
Massey University, Private Bag 102-904 NSMC, Auckland, New Zealand.
}

\date{Received: date / Accepted: date}

\maketitle

\begin{abstract}
We consider the effects of correlations between the in- and out-degrees of individual
neurons on the dynamics of a network of neurons. By using theta neurons, we can derive a set of
coupled differential equations for the expected dynamics of neurons with the same in-degree.
A Gaussian copula is used to introduce correlations between a neuron's in- and out-degree
and numerical bifurcation analysis is used determine the effects of these correlations
on the network's dynamics. For excitatory coupling we find that inducing positive correlations
has a similar effect to increasing the coupling strength between neurons, 
while for inhibitory coupling it has the opposite effect.
We also determine the propensity of various two- and three-neuron
motifs to occur as correlations are varied and give a plausible explanation for the
observed changes in dynamics. 
\keywords{degree correlations \and copula \and theta neuron \and Ott/Antonsen}
\end{abstract}

\section{Introduction}
Determining the effects of a network's structure on its dynamics is an issue of great
interest, particularly in the case of a network of 
neurons~\cite{rox11,schkih15,nykfri17,marhou16}. Since neurons form {\em directed} synaptic
connections, a neuron has both an in-degree --- the number of neurons connected to it,
and an out-degree --- the number of neurons it connects to. In this paper we present a framework
for investigating the effects of correlations, both positive and negative, between these two
quantities. To isolate the effects of these correlations we assume no other structure in the
networks, i.e.~random connectivity based on the neurons' degrees.

A number of other authors have considered this issue and we now summarise relevant 
aspects of their 
 results. LaMar and Smith~\cite{lamsmi10} considered directed
networks of identical pulse-coupled 
phase oscillators and mostly concentrated on the probability that the network would
fully synchronise, and the time taken to do so.
Vasquez et al.~\cite{vashou12} considered binary neurons whose states were updated at discrete
times, and found that negative degree correlations stabilised a low firing rate state,
for excitatory coupling. A later paper~\cite{marhou16} considered more realistic spiking
neurons, had a mix of excitatory and inhibitory neurons, and concentrated more on the
network's response to transient stimuli, as well as analysis of network properties such as mean
shortest path. Several authors have considered networks for which the in- and out-degrees
of a neuron are equal, thereby inducing positive correlations between them~\cite{schkih15,kahsok17}.

Vegu{\'e} and Roxin~\cite{vegrox19} considered 
large networks of both excitatory and inhibitory leaky integrate-and-fire
neurons and used a mean-field formalism to determine steady state distributions of firing
rates within neural populations. They considered the effects of within-neuron degree correlations
for the excitatory to excitatory connections, and sometimes varied the probability of
inhibitory to excitatory connections in order to create a ``balanced state''.
Nykamp et al.~\cite{nykfri17} also considered large networks of 
both excitatory and inhibitory neurons
and used a Wilson-Cowan type firing rate model to investigate the effects of 
within-neuron degree correlations. They showed that once correlations were included,
the dynamics are effectively four-dimensional, in contrast to the two-dimensional
dynamics expected from a standard rate-based excitatory/inhibitory network. They
also related the degree distributions to cortical motifs.
Experimental evidence for within-neuron degree correlations is given in~\cite{Vegper17}.

The structure of the paper is as follows. In Sec.~\ref{sec:model} we present the model
network and summarise the analysis of~\cite{chahat17} showing that under certain
assumptions, the network can be described by a coupled set of ordinary differential equations,
one for the dynamics associated with each distinct in-degree. In Sec.~\ref{sec:cop} we discuss how
to generate correlated in- and out-degrees using a Gaussian copula. Our model
involves sums over all distinct in-degrees, and in Sec.~\ref{sec:red} we present a computationally
efficient method for evaluating these sums, in analogy with Gaussian quadrature.
Our main results are in Sec.~\ref{sec:res} and we show in Sec.~\ref{sec:ML} that they also
occur in networks of more realistic Morris-Lecar spiking neurons. We discuss motifs
in Sec.~\ref{sec:motif} and conclude in Sec.~\ref{sec:conc}.

\section{Model}
\label{sec:model}
We consider the same model of pulse-coupled theta neurons as in~\cite{chahat17}.
The governing equations are
\be
   \frac{d\theta_i}{dt}=1-\cos{\theta_i}+(1+\cos{\theta_i})(\eta_i+I_i) \label{eq:dtheta}
\ee
for $i=1,2\dots N$, where the phase angle $\theta_i$ characterises the state of neuron $i$,
which fires an action potential as $\theta_i$ increases through $\pi$,
\be
   I_i=\frac{K}{\mk}\sum_{j=1}^N A_{ij}P_n(\theta_j),
\ee
$K$ is the strength of connections within the network, $A_{ij}=1$ if there is a connection
from neuron $j$ to neuron $i$ and $A_{ij}=0$ otherwise, $\mk$ is the average 
degree,~$\sum_{i,j}A_{ij}/N$, and $P_n(\theta)=a_n(1-\cos{\theta})^n$ where $a_n$ is
chosen such that $\int_0^{2\pi}P_n(\theta)d\theta=1$. The function $P_n(\theta_j)$ models
the pulse of current emitted by neuron $j$ when it fires and can be made arbitrarily 
``spike-like'' and localised around $\theta_j=\pi$ by increasing $n$.
The parameter $\eta_i$ is the 
input current to neuron $i$ in the absence of coupling and the $\eta_i$
are independently
and randomly chosen from a Lorentzian distribution
\be
   g(\eta)=\frac{\Delta/\pi}{(\eta-\eta_0)^2+\Delta^2} \label{eq:lor}
\ee
Chandra et al.~\cite{chahat17} considered the limit of large $N$ and assumed that 
 the network
can be characterised by two functions.  Firstly
a degree distribution $P(\bk)$, normalised so that $\sum_\bk P(\bk)=N$,
where $\bk=(k_{in},k_{out})$ and $k_{in}$ and $k_{out}$ are the in- and out-degrees, respectively
of a neuron with degree $\bk$. Secondly, an assortativity function $a(\bk' \rightarrow \bk)$
giving the probability of a connection from a neuron with degree $\bk'$ to one with degree
$\bk$, given that such neurons exist. Whereas~\cite{chahat17} investigated the effects of
varying $a(\bk' \rightarrow \bk)$, here we consider the default value for this function
(i.e.~its value expected by chance, see~\eqref{eq:a})
and investigate the effects of varying correlations between $k_{in}$ and $k_{out}$
as specified by the degree distribution $P(\bk)$. We emphasise that we are only considering
{\em within-neuron} degree correlations and are not considering degree assortativity,
which refers to the probability of neurons with specified degrees being connected to one
another~\cite{chahat17,resott14}.

In the limit $N\rightarrow\infty$, the network can be described by a probability
distribution $f(\theta,\eta|\bk,t)$, where $f(\theta,\eta|\bk,t)d\theta\ d\eta$
is the probability that a neuron with degree $\bk$
has phase angle in $[\theta,\theta+d\theta]$ and
value of $\eta$ in $[\eta,\eta+d\eta]$ at time $t$. This distribution satisfies the
continuity equation
\be
   \frac{\partial f}{\partial t}+\frac{\partial}{\partial\theta}(vf)=0 \label{eq:cont}
\ee
where $v$ is the continuum version of the right hand side of~\eqref{eq:dtheta}:
\begin{align}
   v(\theta,\bk,\eta,t) & =  1-\cos{\theta}+(1+\cos{\theta}) \nonumber \\
 & \left[\eta+\frac{K}{\mk}\sum_{\bk'}P(\bk')a(\bk' \rightarrow \bk) \right. \nonumber \\
 & \left. \times \int_{-\infty}^\infty\int_0^{2\pi} f(\theta',\eta'|\bk',t) P_n(\theta')d\theta'\ d\eta'\right] \label{eq:v}
\end{align}
The system~\eqref{eq:cont}-\eqref{eq:v} is amenable to the use of the Ott/Ant\-on\-sen 
ansatz~\cite{ottant08,ottant09} and using standard techniques~\cite{lukbar13,lai14A,lai16,coobyr19}
one can show that the long-time dynamics of the system is described by
\begin{align}
   \frac{\partial b({\bf k},t)}{\partial t} & =\frac{-i(b({\bf k},t)-1)^2}{2}+\frac{(b({\bf k},t)+1)^2}{2}\Bigg[-\Delta  \nonumber \\
 & \left. +i\eta_0+\frac{iK}{\langle k\rangle}
\sum_{{\bf k}'}P({\bf k}') a({\bf k}'\rightarrow {\bf k})G(\mathbf{k}',t) \right] \label{eq:dbdt}
\end{align}
where (having chosen $n=2$)
\begin{gather}
   G(\mathbf{k}',t) \nonumber \\
  =1-\frac{2(b({\bf k}',t)+\bar{b}({\bf k}',t))}{3}+\frac{b({\bf k}',t)^2+\bar{b}({\bf k}',t)^2}{6}.
\end{gather}

The quantity
\be
   b({\bf k},t)=\int_{-\infty}^\infty\int_0^{2\pi} f(\theta,\eta|\bk,t)e^{i\theta}d\theta\ d\eta
\ee
can be regarded as a complex-valued ``order parameter'' for neurons with degree $\bk$ at time $t$.
The function $G(\mathbf{k}',t)$ can be regarded as the output current from neurons
with degree $\bk'$, and its form results from rewriting the pulse function $P_n(\theta)$
in terms of $b(\mathbf{k}',t)$. [For general $n$, $G(\mathbf{k}',t)$ is the sum of a degree-$n$ polynomial in $b(\mathbf{k}',t)$ and 
one in $\bar{b}(\mathbf{k}',t)$ (the conjugate of $b(\mathbf{k}',t)$)~\cite{lai14A,lukbar13}.
One can take the limit $n\rightarrow\infty$ and obtain 
$G(\mathbf{k}',t)=(1-|b(\mathbf{k}',t)|^2)/(1+b(\mathbf{k}',t)+\bar{b}(\mathbf{k}',t)+|b(\mathbf{k}',t)|^2)$.]
Note that the parameters of the Lorenztian~\eqref{eq:lor} appear in~\eqref{eq:dbdt}
as a result of evaluating the integral over $\eta'$ in~\eqref{eq:v}. 
The equation~\eqref{eq:dbdt} only describes the long-time asymptotic behaviour of the 
network~\eqref{eq:dtheta}, on the ``Ott/Antonsen manifold'', and thus may not fully describe
transients from arbitrary initial conditions, nor the effects of stimuli which move the network
off this manifold.

One can also
marginalise $f(\theta,\eta|\bk,t)$ over $\eta$ to obtain
the distribution of $\theta$ for each $\bk$ and $t$:
\begin{gather}
    p_{\theta}(\theta|\bk,t) \nonumber \\
=\frac{1-|b({\bf k},t)|^2}{2\pi\{1-2|b({\bf k},t)|\cos{[\theta-\arg(b({\bf k},t))]}+|b({\bf k},t)|^2\}}
\end{gather}
a unimodal function with maximum at $\theta=\arg(b({\bf k},t))$. The firing rate of neurons
with degree $\bk$ is equal to the flux through $\theta=\pi$, i.e.
\begin{align}
   f({\bf k},t) & =2p_{\theta}(\pi|\bk,t) \nonumber \\
   & =\frac{1-|b({\bf k},t)|^2}{\pi\{1+2|b({\bf k},t)|\cos{[\arg(b({\bf k},t))]}+|b({\bf k},t)|^2\}} \nonumber \\
  & = \frac{1}{\pi}\mbox{Re}\left(\frac{1-\bar{b}({\bf k},t)}{1+\bar{b}({\bf k},t)}\right)
\end{align}
where we have used the fact that $d\theta/dt=2$ when $\theta=\pi$.

Suppose our network has neutral assortativity, i.e.~neurons
are randomly connected with the probability of connection being determined by just their
relevant degrees. Then~\cite{resott14,chahat17}
\be
    a({\bf k}'\rightarrow {\bf k})=\frac{k_{out}'k_{in}}{N\langle k \rangle} \label{eq:a}
\ee
and (writing $P(k_{in}',k_{out}',\hat{\rho})$ instead of $P(\bk')$ from now on, where
$\hat{\rho}$ is a parameter used to calibrate the 
desired correlation between $k_{in}'$ and $k_{out}'$, defined below in~\eqref{eq:biv}) 
\begin{gather}
    \sum_{k_{in}'}\sum_{k_{out}'}P(k_{in}',k_{out}',\hat{\rho})a({\bf k}'\rightarrow {\bf k})G(k_{in}',k_{out}',t) \nonumber \\
  = \frac{k_{in}}{N\langle k\rangle}\sum_{k_{in}'}\sum_{k_{out}'}P(k_{in}',k_{out}',\hat{\rho})k_{out}'G(k_{in}',k_{out}',t) \label{eq:inp}
\end{gather}
This quantity is proportional to the input to a neuron with degree 
$(k_{in},k_{out})$ from other neurons within the network but it is clearly
independent of $k_{out}$, so
the state of a neuron with degree $(k_{in},k_{out})$
must also be independent of $k_{out}$, and thus $G$ must be
independent of $k_{out}'$. So the expression in~\eqref{eq:inp} can be written
\begin{gather}
  \frac{k_{in}}{N\langle k\rangle}\sum_{k_{in}'}Q(k_{in}',\hat{\rho})G(k_{in}',t)
\end{gather}
where
\be
   Q(k_{in}',\hat{\rho})\equiv\sum_{k_{out}'}P(k_{in}',k_{out}',\hat{\rho})k_{out}' \label{eq:Qa}
\ee
The function $Q$ can be thought of as a $k_{in}'$-dependent mean of $k_{out}'$ which
is also dependent on
the correlations between $k_{in}'$ and $k_{out}'$.


Our model equations are thus
\begin{gather}
   \frac{\partial b(k_{in},t)}{\partial t}  =\frac{-i(b(k_{in},t)-1)^2}{2}+\frac{(b(k_{in},t)+1)^2}{2}  \nonumber \\
    \times \left[-\Delta+i\eta_0+\frac{iK k_{in}}{N\langle k\rangle^2}
\sum_{k_{in}'}Q(k_{in}',\hat{\rho})G(k_{in}',t) \right] \label{eq:dbdtA}
\end{gather}
where $k_{in}$ takes on integer values between the minimum and maximum in-degrees.
The correlation between in- and out-degrees of a neuron is controlled by $\hat{\rho}$, 
as explained below, and this
appears as a parameter in~\eqref{eq:Qa}.

It is interesting to compare~\eqref{eq:Qa}-\eqref{eq:dbdtA} with the heuristic rate equation
in~\cite{nykfri17}. These authors characterised a neuron by its ``f-I curve'' --- a nonlinear
function transforming input current into a firing rate. They concluded that the input current
to a neuron is proportional to two quantities: (i) its in-degree, and 
(ii) the sum over in- and out-degrees
of presynaptic neurons of the product of the joint degree distribution, the
out-degree of the presynaptic neuron, and the ``output'' of presynaptic neurons. We also
find this form of equation.

We note that the transformation $V=\tan{(\theta/2)}$ maps a theta neuron to a quadratic
integrate-and-fire (QIF) neuron with threshold and resets of $\pm\infty$, and that for the special
case $n=\infty$ one could derive an equivalent pair of real equations rather than the single
equation~\eqref{eq:dbdtA} where the two real variables are the mean voltage and
firing rate of the QIF
neurons with a specific in-degree~\cite{monpaz15}.

\section{Generating correlated in- and out-degrees}
\label{sec:cop}
We now turn to the problem of deriving $P(k_{in}',k_{out}',\hat{\rho})$ and thus 
$Q(k_{in}',\hat{\rho})$. For simplicity we choose the distributions of both the
in- and out-degrees to be the same, namely power law distributions with exponent
$-3$, truncated below and above at degrees $a$ and $b$ respectively. 
(Evidence for power law distributions in the human
brain is given in~\cite{eguchi05}, for example.)
So the probability distribution function of either in- or out-degree $k$ is
\be
	   p(k)=\begin{cases} \left(\frac{2a^2b^2}{b^2-a^2}\right)k^{-3} & a\leq k\leq b \\ 0 & \mbox{ otherwise} \end{cases}
\ee
where the normalisation factor results from approximating the sum from $a$ to $b$ by an integral.
(The approximation improves as $a$ and $b$ are both increased.)
We want to introduce correlations between the in- and out-degree of a neuron, while retaining
these marginal distributions. We do this using a Gaussian copula~\cite{nel07}. The correlated
bivariate normal distribution with zero mean is
\begin{align}
   f(x,y,\hat{\rho}) & =\frac{1}{2\pi\sqrt{|\Sigma|}}e^{-(\mathbf{x}^T\Sigma^{-1}\mathbf{x})/2} \nonumber \\
 & =\frac{1}{2\pi\sqrt{1-\hat{\rho}^2}}e^{-(x^2-2\hat{\rho} xy+y^2)/[2(1-\hat{\rho}^2)]} \label{eq:biv}
\end{align}
where 
\be
   \mathbf{x}\equiv\begin{pmatrix} x \\ y \end{pmatrix} \qquad 
\Sigma=\begin{pmatrix} 1 & \hat{\rho} \\ \hat{\rho} & 1 \end{pmatrix}
\ee
and $\hat{\rho}\in(-1,1)$ is the correlation between $x$ and $y$. 
The variables $x$ and $y$ have no physical meaning and we use the copula just as a way of
deriving an analytic expression for $P(k_{in}',k_{out}',\hat{\rho})$ for which the
correlations between $k_{in}'$ and $k_{out}'$ can be varied systematically.

The marginal distributions for $x$ and $y$ are the same:
\be
   \tilde{p}(x)=\frac{1}{\sqrt{2\pi}}e^{-x^2/2}
\ee
as are their cumulative distribution functions:
\be
   C(x)=[1+\mbox{erf}(x/\sqrt{2})]/2
\ee
We define the cumulative distribution function of $f$:
\be
   F(X,Y,\hat{\rho})=\int_{-\infty}^Y\int_{-\infty}^X f(x,y,\hat{\rho})dx\ dy
\ee
and also have the cumulative distribution function for a degree $k$:
\be
   C_k(k)= \int_a^k \left(\frac{2a^2b^2}{b^2-a^2}\right)s^{-3}ds=\frac{b^2(k^2-a^2)}{k^2(b^2-a^2)}
\ee
where we have treated $k$ as a continuous variable and again approximated a sum by an integral.

\begin{figure}
\includegraphics[scale=0.45]{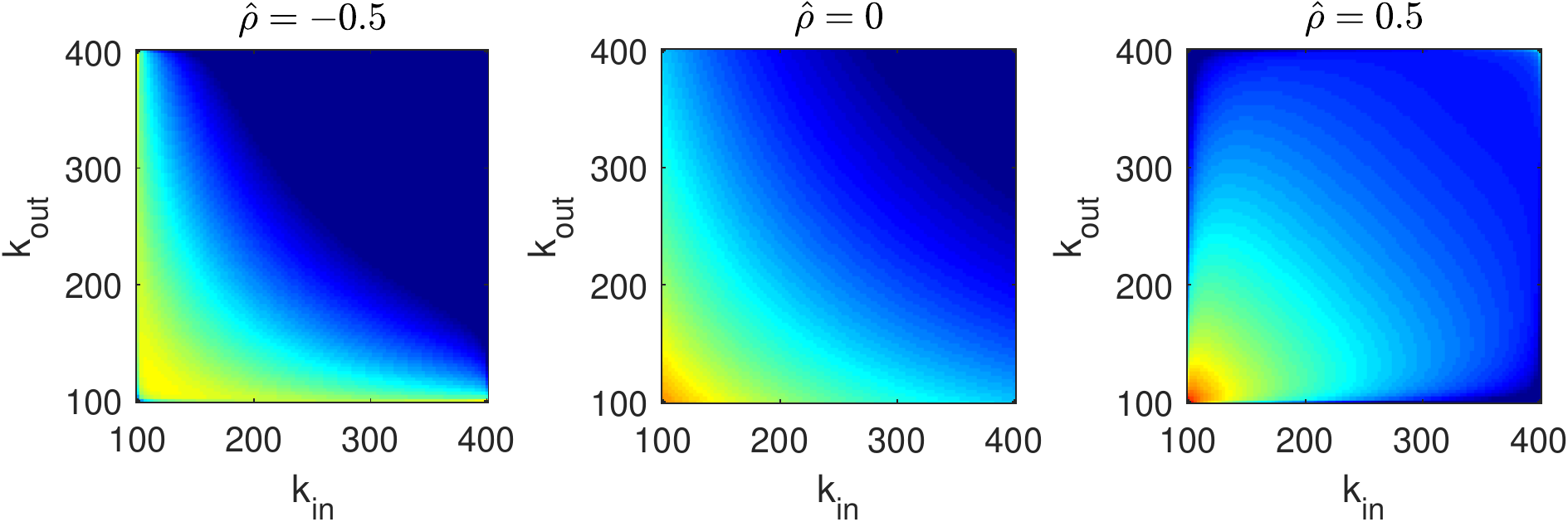}
\caption{Log of $P(k_{in},k_{out},\hat{\rho})$ is shown for three different values
of $\hat{\rho}$ (red: larger $P$, blue: smaller $P$). $a=100,b=400$.}
\label{fig:copula}
\end{figure}

We thus have the joint cumulative distribution function for $k_{in}$ and $k_{out}$
\begin{gather}
   \widehat{C}(k_{in},k_{out},\hat{\rho})  =F(C^{-1}(C_k(k_{in})),C^{-1}(C_k(k_{out})),\hat{\rho}) \nonumber \\
  =\int_{-\infty}^{C^{-1}(C_k(k_{out}))}\int_{-\infty}^{C^{-1}(C_k(k_{in}))} f(x,y,\hat{\rho})dx\ dy
\end{gather}
The joint degree distribution for  $k_{in}$ and $k_{out}$ is then
\begin{gather}
   P(k_{in},k_{out},\hat{\rho})  =  \frac{\partial^2}{\partial k_{in}\partial k_{out}}\widehat{C}(k_{in},k_{out},\hat{\rho}) \nonumber \\
  =  \{C^{-1}[C_k(k_{in})]\}'\{C^{-1}[C_k(k_{out})]\}'\nonumber \\
\times f\{C^{-1}[C_k(k_{in})],C^{-1}[C_k(k_{out}]),\hat{\rho}\} \label{eq:jointk}
\end{gather}
where the primes indicate differentiation with respect to the relevant $k$. Now
\be
   C^{-1}(x)=\sqrt{2}\mbox{ erf}^{-1}(2x-1)
\ee
so
\be
   C^{-1}[C_k(k)]= \sqrt{2}\mbox{ erf}^{-1}\left(\frac{2b^2(k^2-a^2)}{k^2(b^2-a^2)}-1\right)
\ee
and
\begin{gather}
   \{C^{-1}[C_k(k)]\}'= \nonumber \\
\sqrt{\frac{\pi}{2}}\exp{\left[\left\{\mbox{erf}^{-1}\left(\frac{2b^2(k^2-a^2)}{k^2(b^2-a^2)}-1\right)\right\}^2\right]}\frac{4a^2b^2}{(b^2-a^2)k^3}
\end{gather}
Substituting these into~\eqref{eq:jointk} and simplifying we find
\begin{gather}
   P(k_{in},k_{out},\hat{\rho})  =\frac{4a^4b^4}{\sqrt{1-\hat{\rho}^2}(b^2-a^2)^2k_{in}^3k_{out}^3} \nonumber \\
 \times\exp{\left\{\frac{\hat{\rho}C^{-1}[C_k(k_{in})]C^{-1}[C_k(k_{out})}{1-\hat{\rho}^2}\right\}} \nonumber \\
  \times \exp{\left[\frac{-\hat{\rho}^2\left(\{C^{-1}[C_k(k_{in})]\}^2+\left\{C^{-1}[C_k(k_{out})]\right\}^2\right)}{2(1-\hat{\rho}^2)}\right]}  \\
    = \frac{p(k_{in})p(k_{out})}{\sqrt{1-\hat{\rho}^2}}\exp{\left\{\frac{\hat{\rho}C^{-1}[C_k(k_{in})]C^{-1}[C_k(k_{out})]}{1-\hat{\rho}^2}\right\}} \nonumber \\
   \times \exp{\left[\frac{-\hat{\rho}^2\left(\{C^{-1}[C_k(k_{in})]\}^2+\left\{C^{-1}[C_k(k_{out})]\right\}^2\right)}{2(1-\hat{\rho}^2)}\right]} \label{eq:cop}
\end{gather}
Note that for $\hat{\rho}=0$, this simplifies to $p(k_{in})p(k_{out})$, as expected.
Examples of $P(k_{in},k_{out},\hat{\rho})$ for different $\hat{\rho}$ are shown
in Fig.~\ref{fig:copula}.
Both Zhao et al.~\cite{zhabev11} and LaMar and Smith~\cite{lamsmi10}
used Gaussian copulas to create networks with correlated
in- and out-degrees as done here, but did not derive an expression of the form~\eqref{eq:cop}. 

\begin{figure}
\includegraphics[scale=0.65]{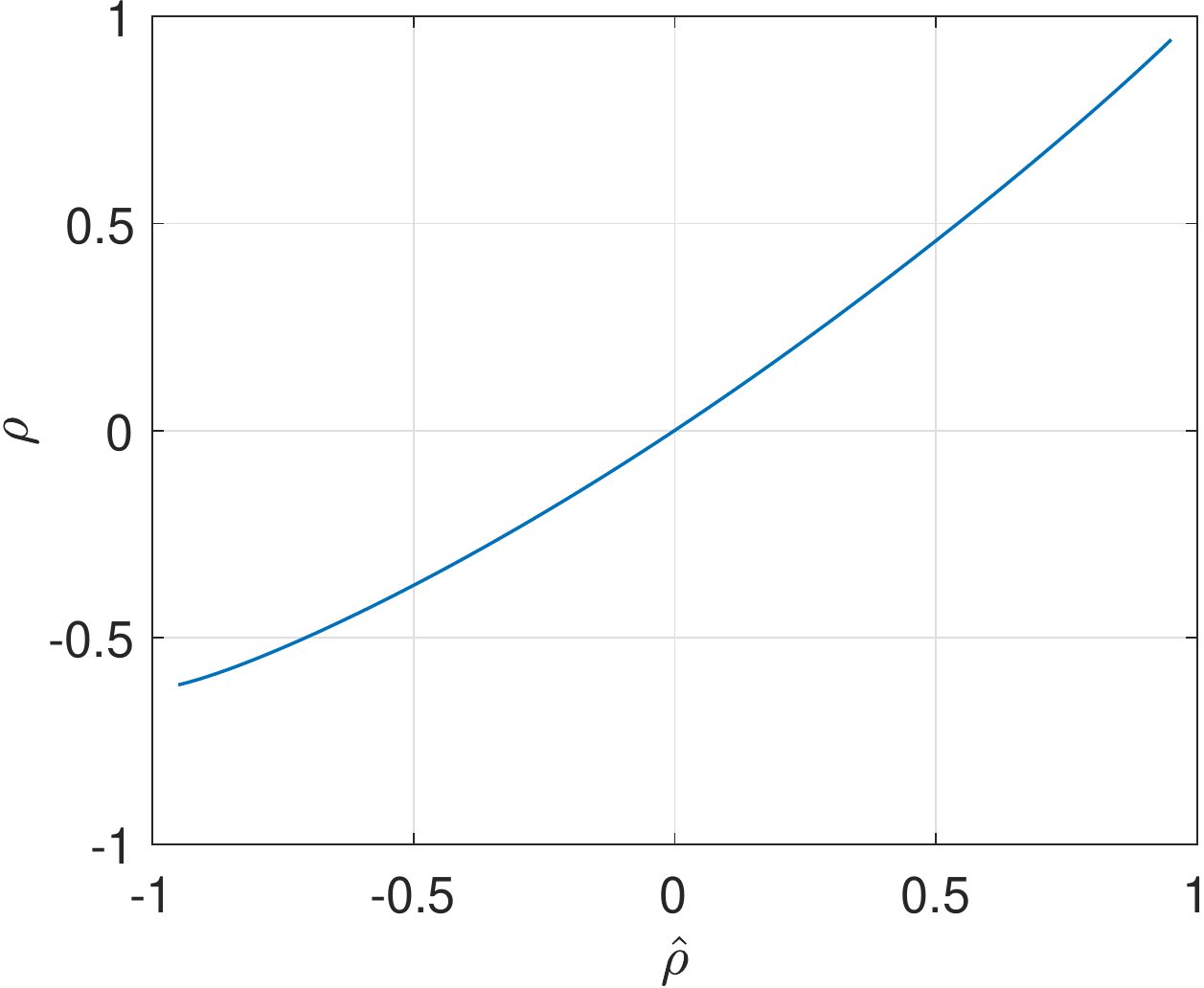}
\caption{Correlation coefficient between in- and out-degrees, $\rho$,
as a function of the correlation coefficient in the Gaussian copula, $\hat{\rho}$.
Parameters:  $a=100,b=400$.}
\label{fig:rhorho}
\end{figure}

We need to relate $\hat{\rho}$, a parameter in~\eqref{eq:cop}, 
to $\rho$, the Pearson's correlation coefficient between in- and
out-degrees of a neuron (note: not between two connected neurons). We have
\be
   \rho=\frac{\tilde{\Sigma}P(k_{in},k_{out},\hat{\rho})(k_{in}-\mk)(k_{out}-\mk)}
{\sqrt{\tilde{\Sigma}P(k_{in},k_{out},\hat{\rho})(k_{in}-\mk)^2}\sqrt{\tilde{\Sigma}P(k_{in},k_{out},\hat{\rho})(k_{out}-\mk)^2}}
\ee
where $\tilde{\Sigma}$ indicates a sum over all $k_{in}$ and $k_{out}$.
$\rho$ as a function of $\hat{\rho}$ is shown in 
Fig.~\ref{fig:rhorho}. We see that the relationship is monotonic, and while it is possible
to obtain values of $\rho$ close to 1, the lower limit is approximately $-0.6$. 
By varying $\hat{\rho}$ in~\eqref{eq:dbdtA} we can thus investigate the effects of 
varying the correlation coefficient between in- and
out-degrees of a neuron ($\rho$) on the dynamics of a network.
Note that for the distributions used here, treating $k$ as a continuous variable, 
$\langle k \rangle=2ab/(b+a)$. 

\begin{figure}
\includegraphics[scale=0.4]{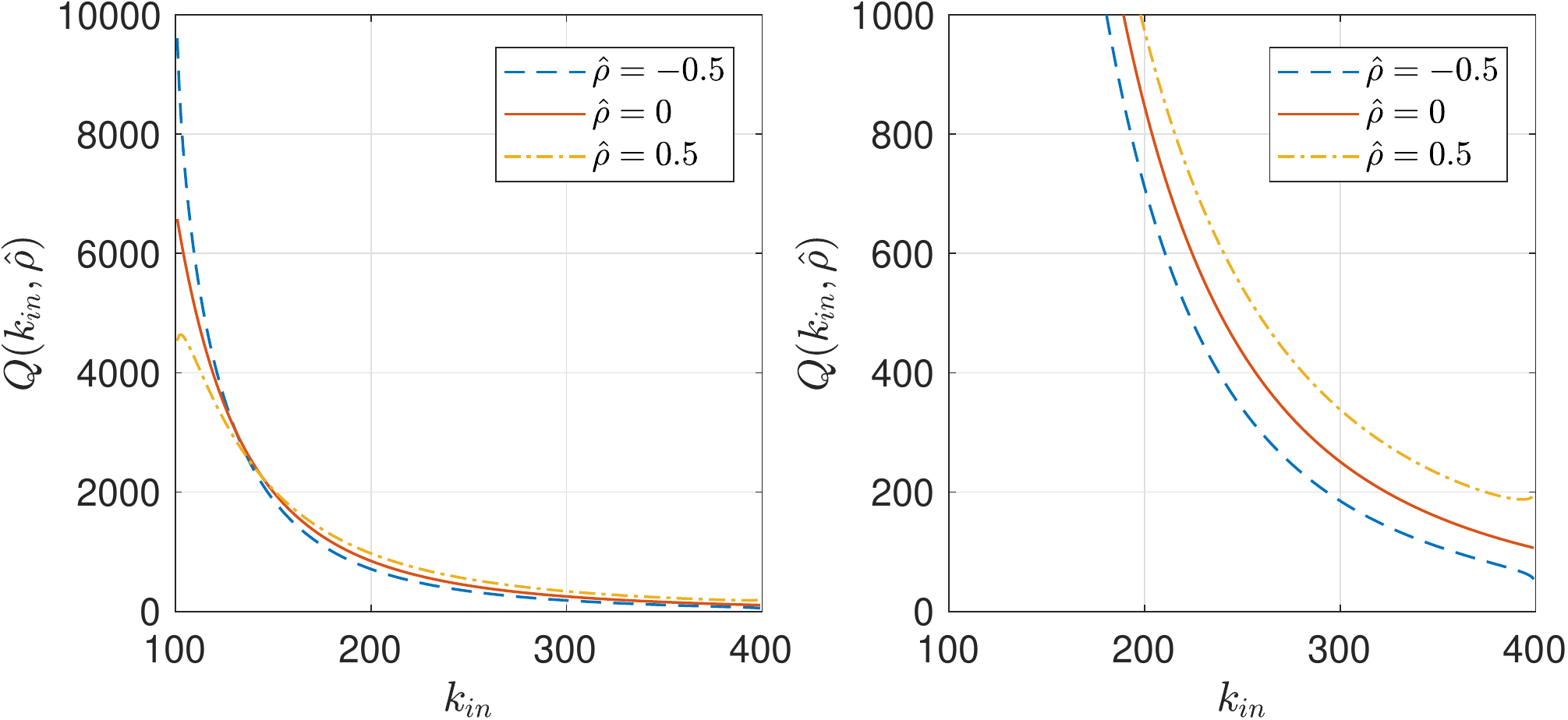}
\caption{The function $Q(k_{in},\hat{\rho})$ (eqn.~\eqref{eq:Q}) 
for different $\hat{\rho}$. The right panel is a zoom of the left panel.
Parameters: $a=100,b=400,N=2000$.}
\label{fig:ptilde}
\end{figure}

Keeping in mind the normalisation $\sum_\bk P(\bk)=N$
we write $Q(k_{in}',\hat{\rho})$ as
\be
   Q(k_{in}',\hat{\rho})=N\sum_{k_{out}'=a}^b P(k_{in}',k_{out}',\hat{\rho})k_{out}' \label{eq:Q}
\ee
Note that the factor of $N$ here cancels with that in the last term in~\eqref{eq:dbdtA},
giving equations which do not explicitly depend on $N$.  
Examples of $Q(k_{in}',\hat{\rho})$ for different $\hat{\rho}$
are shown in Fig.~\ref{fig:ptilde}. We see that increasing
$\hat{\rho}$ gives more weight to high in-degree nodes and less to low in-degree nodes
and vice versa.

\section{Reduced model}
\label{sec:red}

We now turn to the issue of evaluating the sums over degrees in both~\eqref{eq:Q}
and~\eqref{eq:dbdtA}. Although such sums are typically over only several hundred terms,
it is possible to accurately evaluate them using many fewer terms, in analogy with
Gaussian quadrature~\cite{eng06}. 

Defining an inner product as the sum
\be
   (f,g)=\sum_{k=a}^b f(k)g(k)
\ee
we assume that there is a corresponding set of orthogonal polynomials
$\{q_n(k)\}_{0\leq n}$ associated with this
product. These polynomials satisfy the three-term recurrence relationship
\be
   q_{n+1}(k)=(k-\alpha_n)q_n(k)-\beta_n q_{n-1}(k)
\ee
where
\be
   \alpha_n\equiv \frac{(kq_n,q_n)}{(q_n,q_n)}; \qquad 0\leq n
\ee
\be
  \beta_n\equiv \frac{(q_n,q_n)}{(q_{n-1},q_{n-1})}; \qquad 1\leq n
\ee
$q_0(k)=1$ and $q_{-1}(k)=0$. Then for a given positive integer $n$,
assuming that $f$ is $2n$ times continuously differentiable,
 we have the Gaussian summation formula
\be
   \sum_{k=a}^b f(k)=\sum_{i=1}^n w_i f(x_i)+R_n
\ee
with error
\be
   R_n=\frac{f^{(2n)}(\xi)}{(2n)!}(q_n,q_n)
\ee
where $x_i$ are the $n$ roots of $q_n$, $\xi\in[a,b]$, and the weights $w_i$ are discussed below.
Note that the roots of $q_n(k)$ are typically not integers, but this does not matter if the
function $f(k)$ can be evaluated for arbitrary $k$.
 
In practice, to find the roots of $q_n$ we use the Golub-Welsch algorithm. Form the 
tridiagonal matrix
\be
   J=\begin{pmatrix} \alpha_0 & \sqrt{\beta_1} & 0 & \dots & \dots & \dots \\
   \sqrt{\beta_1} & \alpha_1 & \sqrt{\beta_2} & 0 &  \dots & \dots \\
   0 & \sqrt{\beta_2} & \alpha_2 & \sqrt{\beta_3} & 0 & \dots \\
   0 & \dots & \dots & \dots & \dots & 0 \\
   \dots & \dots & 0 & \sqrt{\beta_{n-2}} & \alpha_{n-2} & \sqrt{\beta_{n-1}} \\
  \dots & \dots & \dots & 0 & \sqrt{\beta_{n-1}} & \alpha_{n-1}
\end{pmatrix}
\ee
The eigenvalues of $J$ are the $\{x_i\}$ and if all eigenvectors, $v_i$, are scaled to have norm 1,
then $w_i=(b-a)\left(v_i^{(1)}\right)^2$, where $v_i^{(1)}$ is the first component of $v_i$.

We will use the approximation
\be
   \sum_{k=a}^b f(k)\approx\sum_{i=1}^n w_i f(x_i) \label{eq:sumapprox}
\ee
where $n\ll b-a+1$, the number of terms in the original sum.
Given the resemblence of the sum on the left in~\eqref{eq:sumapprox} to the integral
of $f(k)$ between $k=a$ and $k=b$, it is not surprising that the roots of $p_n$,
when translated from the interval $[a,b]$ to $[-1,1]$, are close to the
roots of the $n$th order Legendre polynomial, as would be used in Gaussian quadrature.
(The same is true for the corresponding weights.)

We thus choose $n$ and write
\be
   Q(k_{in}',\hat{\rho})=N\sum_{j=1}^n w_j P(k_{in}',k_j,\hat{\rho})k_j 
\ee
where $k_j$ are the roots and $w_j$ are the weights, respectively, associated with $q_n(k)$.
In order to use the same approximation for the sum in~\eqref{eq:dbdtA} we consider only
values of $k_{in}$ equal to the $k_j$. As mentioned, these are typically {\em not} integers.
We refer to them as ``virtual degrees''.
Thus our model equations are
\begin{align}
   \frac{\partial b(k_j,t)}{\partial t} & =\frac{-i(b(k_j,t)-1)^2}{2}+\frac{(b(k_j,t)+1)^2}{2}\Bigg[-\Delta \nonumber \\
& \left. +i\eta_0+\frac{iK k_j}{N\langle k\rangle^2}
\sum_{j=1}^n w_j Q(k_j,\hat{\rho})G(k_j,t) \right] \label{eq:dbdtB}
\end{align}
for $j=1,\dots n$. We are interested in fixed points of these equations, and how these
fixed points and their stabilities change as parameters such as $\eta_0$ and $\hat{\rho}$
are varied. We use pseudo-arclength continuation~\cite{lai14B,gov00} to investigate this.

In order to calculate the mean frequency of the network we use the result
that the frequency for neurons with in-degree $k$ is~\cite{monpaz15}
\be
   f(k)=\frac{1}{\pi}\mbox{Re}\left(\frac{1-\bar{b}(k)}{1+\bar{b}(k)}\right),
\ee
where overline indicates complex conjugate,
and then average over the network to obtain the mean frequency
\begin{align}
   f & =\frac{\sum_{k_{in}}\sum_{k_{out}}P(k_{in},k_{out},\hat{\rho})f(k_{in})}{\sum_{k_{in}}\sum_{k_{out}}P(k_{in},k_{out},\hat{\rho})} \nonumber \\
 & =\frac{\sum_{i=1}^n\sum_{j=1}^n w_i w_j P(k_i,k_j,\hat{\rho}) f(k_i)}{\sum_{i=1}^n\sum_{j=1}^n w_i w_j P(k_i,k_j,\hat{\rho})}
\end{align}
(The normalisation is needed because even though the integral of the joint 
degree distribution over $[k_{in},k_{out}]^2$ equals 1, the sum over the
corresponding discrete grid does not.)

\begin{figure}
\includegraphics[scale=0.5]{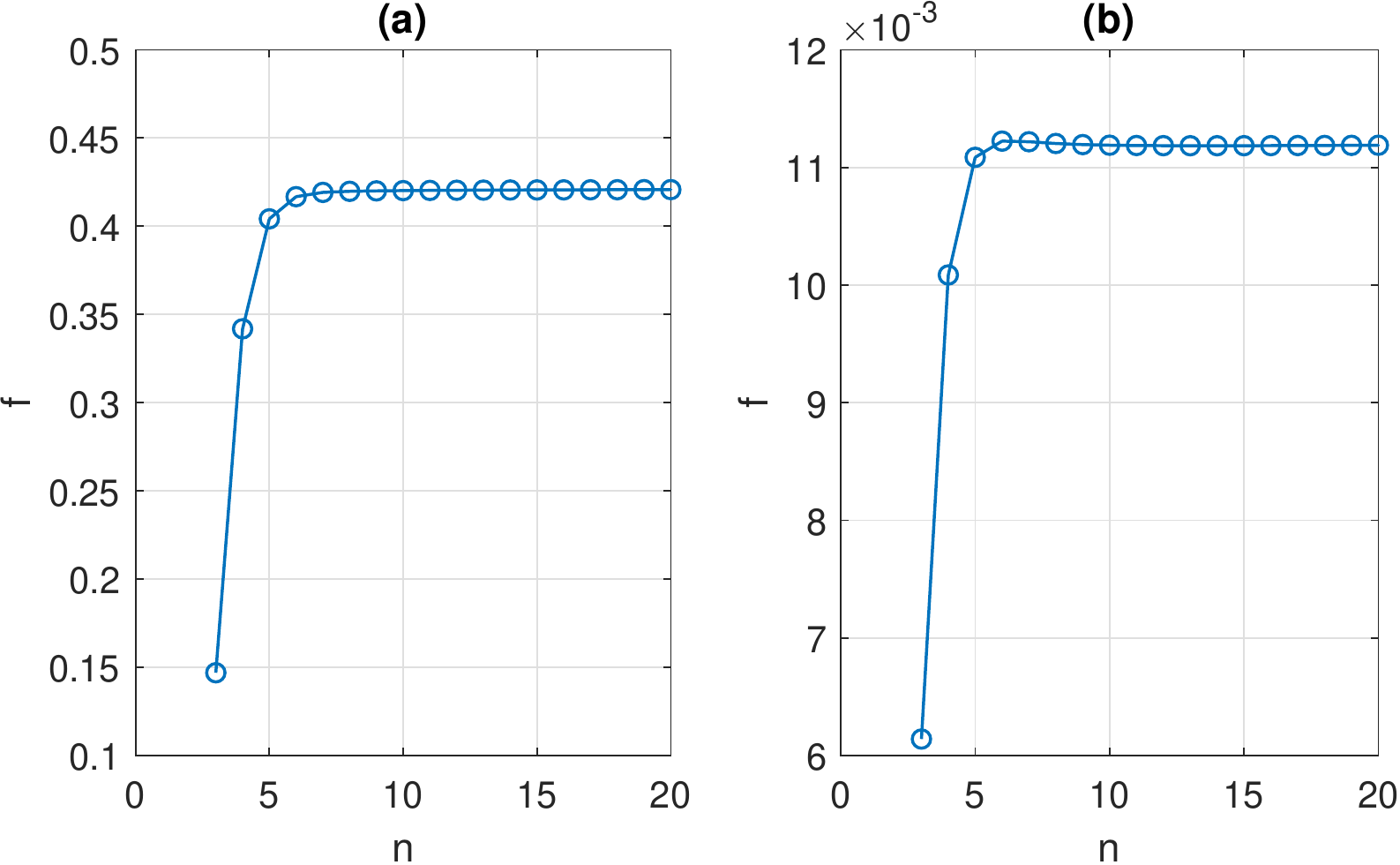}
\caption{Mean frequency, $f$, as a function of $n$, the number of virtual degrees used.
(a): $\hat{\rho}=-0.2,K=1,\eta_0=0.5$. (b):  $\hat{\rho}=0.3,K=-0.1,\eta_0=-0.5$.
Other parameters: $a=100,b=400,\Delta=0.05,N=2000$.}
\label{fig:conv}
\end{figure}

Typical convergence of a calculation of $f$ 
with increasing $n$ is shown in Fig.~\ref{fig:conv} for several sets
of parameter values. We see rapid convergence and  choose $n=15$ for future calculations.
(Calculations of the form shown in Figs.~\ref{fig:bistabA} and~\ref{fig:bistabD} 
were repeated using the full degree sequence from $a$ to $b$, with essentially identical
results.)

\section{Results}
\label{sec:res}


\subsection{Excitatory coupling}

We first consider the case of excitatory coupling, i.e.~$K>0$. We expect a region of bistability
for negative $\eta_0$, as seen in Fig.~\ref{fig:bistabA}. 
We see that
decreasing $\rho$ moves the curve to the right and vice versa. ($\hat{\rho}$ was chosen
to give these particular values of $\rho$.) Following the saddle-node bifurcations as
$\rho$ is varied we obtain Fig.~\ref{fig:snvaryrho}.

\begin{figure}
\includegraphics[scale=0.65]{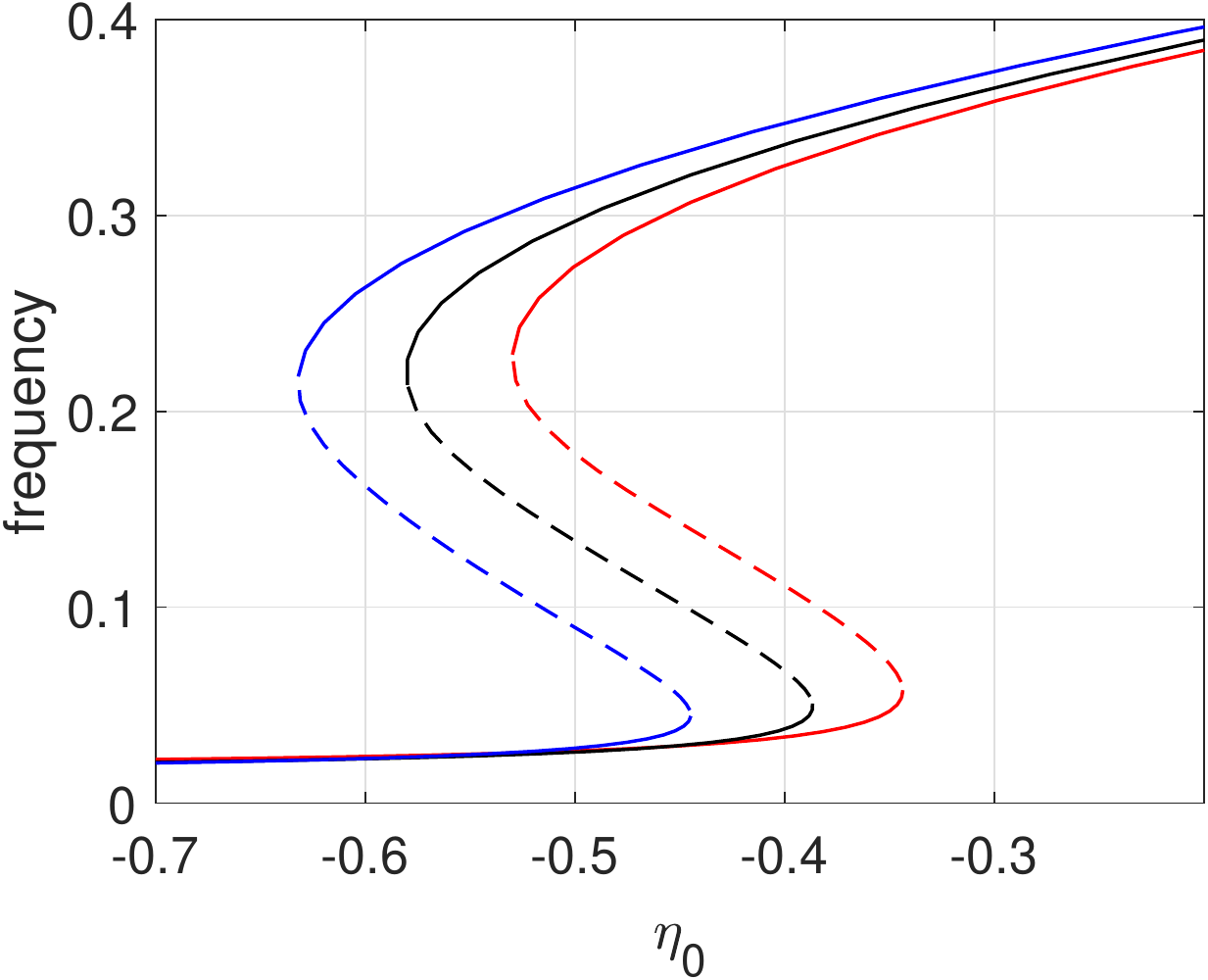}
\caption{Mean frequency, $f$, versus $\eta_0$ for (left to right) $\rho=0.5,0$ and $-0.5$.  
Solid: stable, dashed: unstable. Parameters: $a=100,b=400,K=1.5,\Delta=0.05$.}
\label{fig:bistabA}
\end{figure}

\begin{figure}
\includegraphics[scale=0.65]{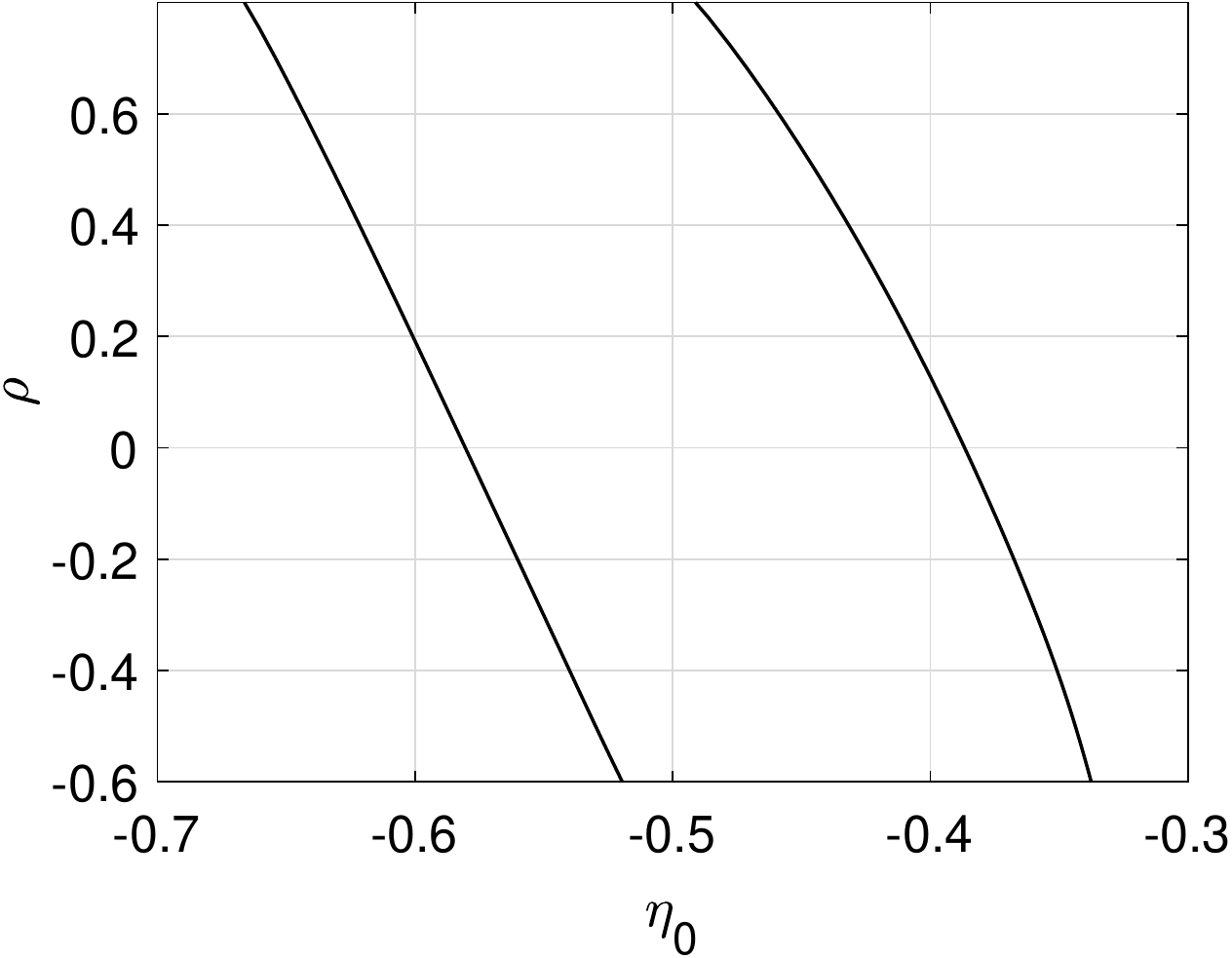}
\caption{Continuation of the saddle-node bifurcations shown in Fig.~\ref{fig:bistabA}.
The network is bistable in the region between the curves.
Parameters as in Fig.~\ref{fig:bistabA}.}
\label{fig:snvaryrho}
\end{figure}

Given the influence of $\hat{\rho}$ (and thus $\rho$) on $Q$ (see Fig.~\ref{fig:ptilde})
this result is easy to understand. Neurons with high in-degree fire faster than those
with low in-degree, and for positive $\rho$, high in-degree neurons contribute more
to the sum in~\eqref{eq:dbdtB} than for negative $\rho$. Thus the total amount of
``output'' from neurons is higher for positive $\rho$ and lower for negative $\rho$.
Put another way, with positive $\rho$, neurons with high firing rate (due to high in-degree)
are more likely to have a high out-degree, thus exciting more neurons than would otherwise be
the case.
Increasing $\rho$ has the same qualitative effect as increasing the coupling strength $K$,
as observed by~\cite{nykfri17}.

\subsection{Inhibitory coupling}
\begin{figure}
\includegraphics[scale=0.55]{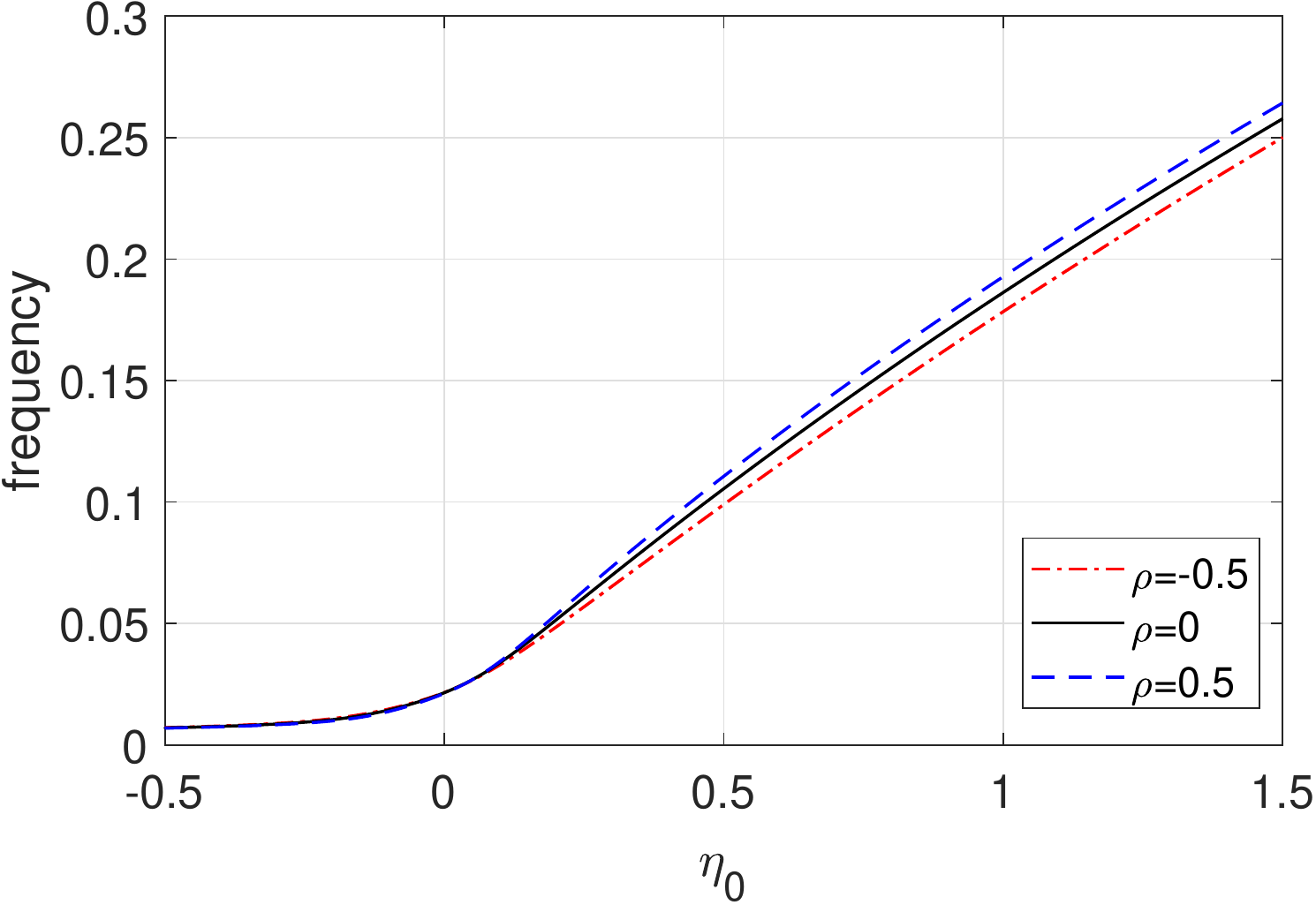}
\caption{Mean frequency, $f$, versus $\eta_0$ for  $\rho=-0.5,0$ and $0.5$;
same colour code as in Fig.~\ref{fig:bistabA}. All branches are stable.
Parameters: $a=100,b=400,K=-1,\Delta=0.05$.}
\label{fig:bistabD}
\end{figure}

Next we consider inhibitory coupling, with $K=-1$. Average network frequency versus $\eta_0$ is shown
in Fig.~\ref{fig:bistabD} for three different values of $\rho$. We see that increasing $\rho$
slightly increases the frequency and vice versa. We can also understand this behaviour
in a qualitative sense. For inhibitory coupling, neurons with high in-degree are not likely
to be firing, so can be ignored. When $\rho<0$, neurons with low in-degree will have high
out-degree, thus the amount of inhibitory ``output'' in the network is increased.
For positive $\rho$, neurons with low in-degree will have low out-degree, thus they will
inhibit fewer neurons than in the case of negative $\rho$, leading to a higher average
firing rate.

We performed calculations corresponding to the results shown in Figs.~\ref{fig:bistabA}
and~\ref{fig:bistabD} for networks of theta neurons and found qualitatively, and to a large
extent quantitatively, the same behaviour as in those figures (results not shown). 


\section{More realistic network}
\label{sec:ML}
To verify the behaviour seen above in a network of theta neurons, we investigated
 a more realistic network of spiking neurons,
in this case Morris-Lecar neurons.
For the case of excitatory coupling the network equations are~\cite{tsukit06}
\begin{align}
   C\frac{dV_i}{dt} & = g_L(V_L-V_i)+g_{Ca}m_\infty(V_i)(V_{Ca}-V_i) \label{eq:dVdt} \\
 & +g_Kn_i(V_K-V_i) \nonumber  \\
  & +I_0+I_i+(V_{ex}-V_i)\frac{\epsilon}{N}\sum_{j=1}^N A_{ij}s_j \nonumber \\
   \frac{dn_i}{dt} & = \frac{\lambda_0(w_\infty(V_i)-n_i)}{\tau_n(V_i)} \\
   \tau\frac{ds_i}{dt} & = m_\infty(V_i)-s_i \label{eq:dsdt}
\end{align}
where
\begin{align}
   m_\infty(V) & =0.5(1+\tanh{[(V-V_1)/V_2]}) \\
   w_\infty(V) & =0.5(1+\tanh{[(V-V_3)/V_4]}) \\
   \tau_n(V) & = \frac{1}{\cosh{[(V-V_3)/(2V_4)]}}
\end{align}
Parameters are $V_1=-1.2,V_2=18,V_3=12,V_4=17.4,\lambda_0=1/15 msec^{-1},g_L=2,
g_K=8,
g_{Ca}=4,
V_L=-60,
V_{Ca}=120,
V_K=-80,
C=20\mu F/cm^2,\tau=100,V_{ex}=120,\epsilon=5 mS/cm^2$. 
Voltages are in mV, conductances are in mS/cm$^2$, time is measured in milliseconds,
and currents in $\mu A/cm^2$.
In the absence of coupling and heterogeneity
a neuron undergoes a SNIC bifurcation as $I_0$ is increased through $\sim 40$.
We have used synaptic coupling of the form in~\cite{ermkop90}, but on a timescale $\tau$ 
rather than instantaneous as in that paper.
The $I_i$ are randomly chosen from a Lorentzian distribution with
mean zero and half-width at half-maximum $0.05$. 

\begin{figure}
\includegraphics[scale=0.45]{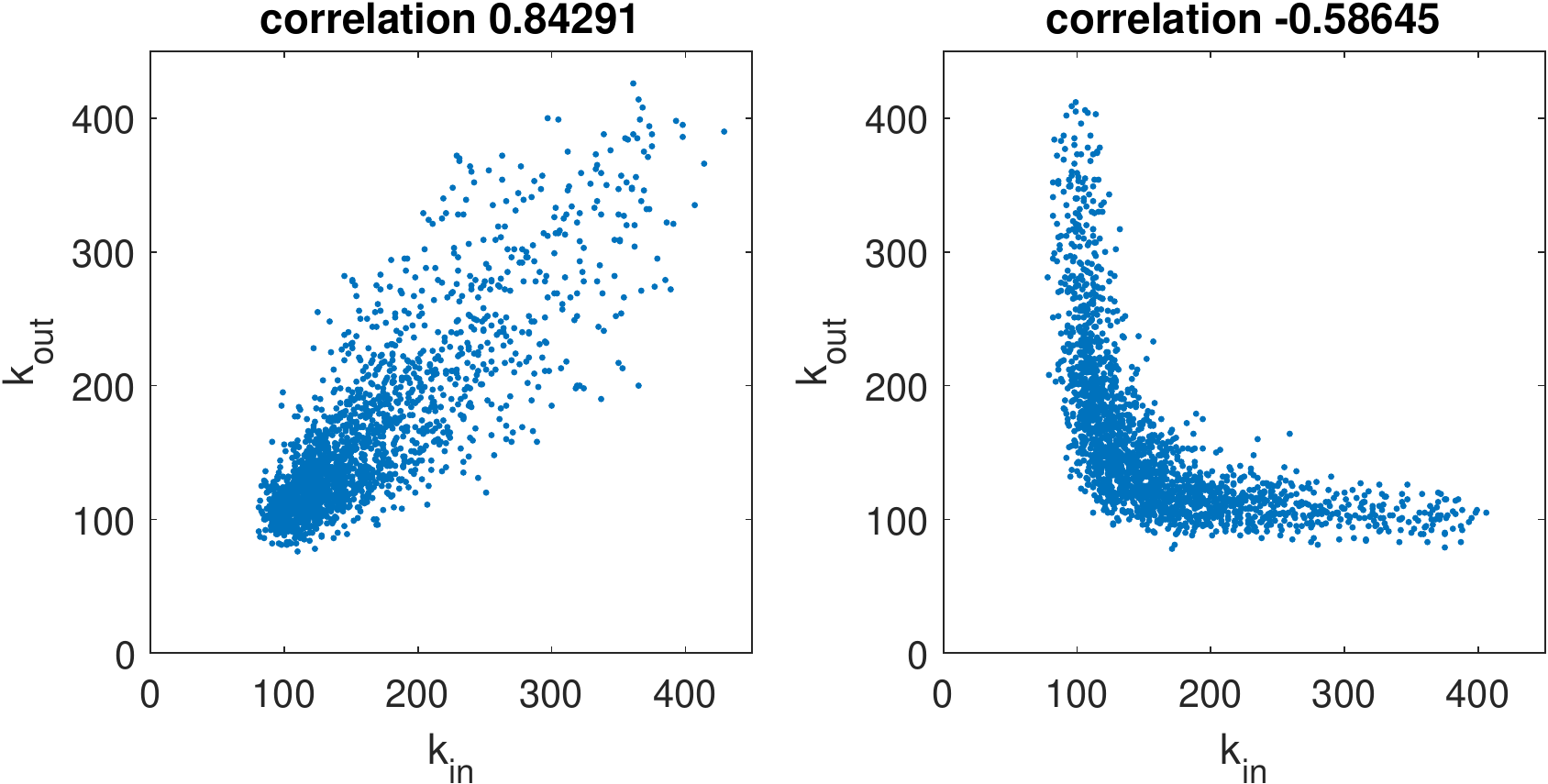}
\caption{Degrees for a network whose generation is described in Sec.~\ref{sec:ML}
for $\hat{\rho}=0.9$ (left) and $\hat{\rho}=-0.9$ (right). Parameters: $N=2000,a=100,b=400$.}
\label{fig:Anet}
\end{figure}

\begin{figure}
\includegraphics[scale=0.45]{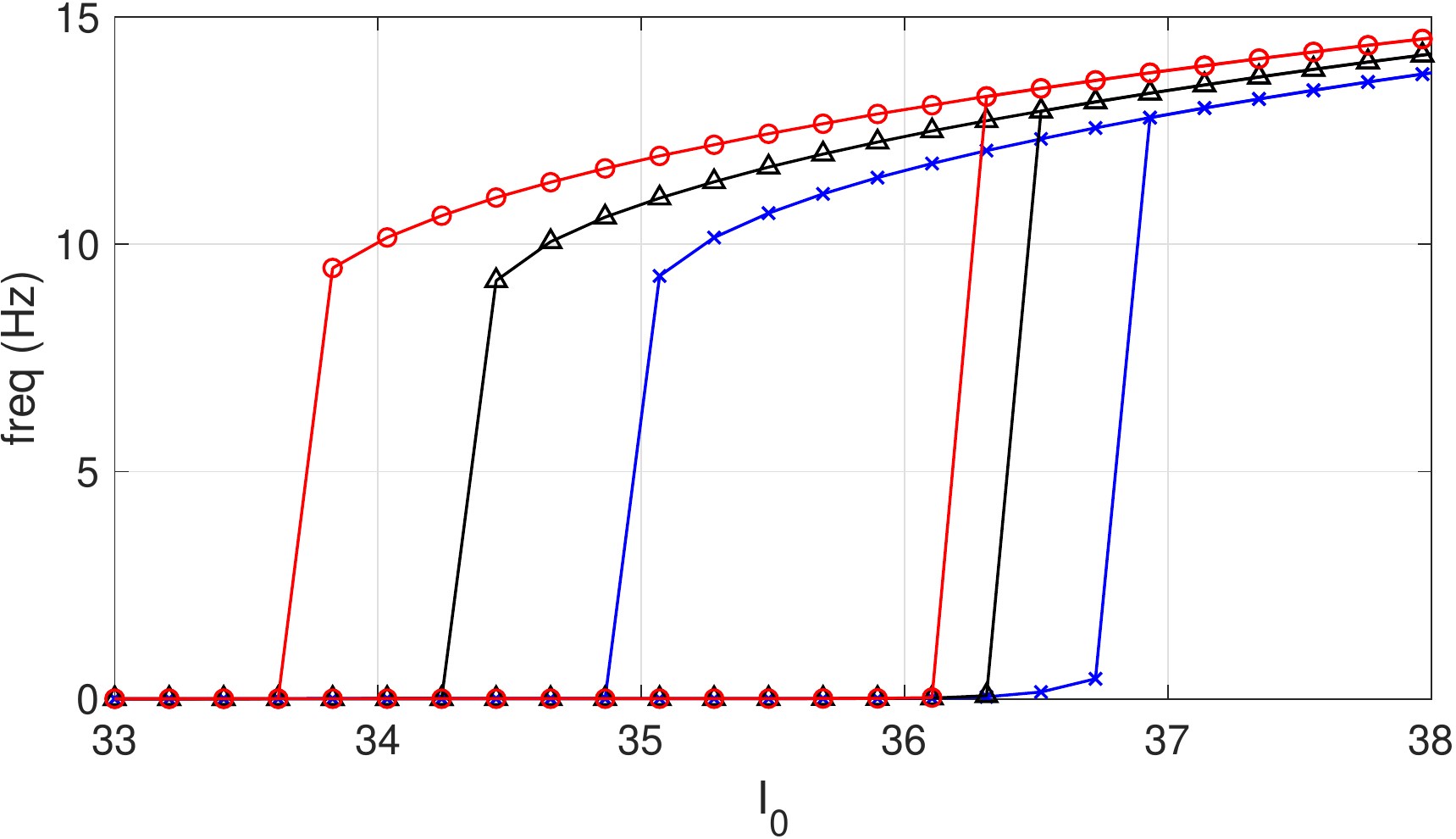}
\caption{Mean frequency versus $I_0$ for a network of Morris-Lecar neurons.  
$N=2000$. Blue crosses: $\rho=-0.57$; black diamonds: $\rho=0$;
red circles: $\rho=0.85$. $I_0$ is quasi-statically increased and then decreased in all
cases.}
\label{fig:bistabB}
\end{figure}

The network is created as follows, using the Gaussian copula of Sec.~\ref{sec:cop}. 
For each $i\in\{1,\dots N\}$
let $x_1$ and $x_2$ be independently chosen from a unit normal distribution. Then $x_1$ and
$y_1=\hat{\rho} x_1+\sqrt{1-\hat{\rho}^2}x_2$ both have unit normal distributions and covariance $\hat{\rho}$, i.e.~are realisations of $x$ and $y$ in~\eqref{eq:biv}.
We then set $k_{in}^i=C_k^{-1}(C(x_1))$ and $k_{out}^i=C_k^{-1}(C(y_1))$.
These degrees each have distribution $p(k)$ but have correlation coefficient $\rho$,
where $\rho$ is determined by the value of $\hat{\rho}$ as shown in Fig.~\ref{fig:rhorho}. 
We then create the connection
from neuron $j$ to neuron $i$ (i.e.~set $A_{ij}=1$) with probability
\be
   \frac{k_{in}^ik_{out}^j}{N\langle k\rangle}
\ee
where $\langle k\rangle$ is the mean of the degrees, and $A_{ij}=0$ otherwise
(the Chung-Lu model~\cite{chulu02}).
Typical results for the network generation
are shown in Fig.~\ref{fig:Anet}, and the measured correlations are 
given in the figure.
The distributions of the resulting
degrees no longer match the distributions of the $k_{in}^i$ and $k_{out}^i$, but are
close. We could have used the configuration model to avoid this problem~\cite{new03},
but here we are only interested in qualitative results.
Quasi-statically sweeping through $I_0$ for networks with three different values of $\rho$ we 
obtain Fig.~\ref{fig:bistabB}, in qualitative
agreement with Fig.~\ref{fig:bistabA}. In Fig.~\ref{fig:bistabA} there is a region of
bistability for each value of $\rho$, and the region moves to lower average drive
as $\rho$ is increased. Since we cannot detect unstable states through simulation 
of~\eqref{eq:dVdt}-\eqref{eq:dsdt}, this bistability is manifested as jumps from low frequency
to high frequency branches as $I_0$ is varied, as seen in Fig.~\ref{fig:bistabB}.

\begin{figure}
\includegraphics[scale=0.45]{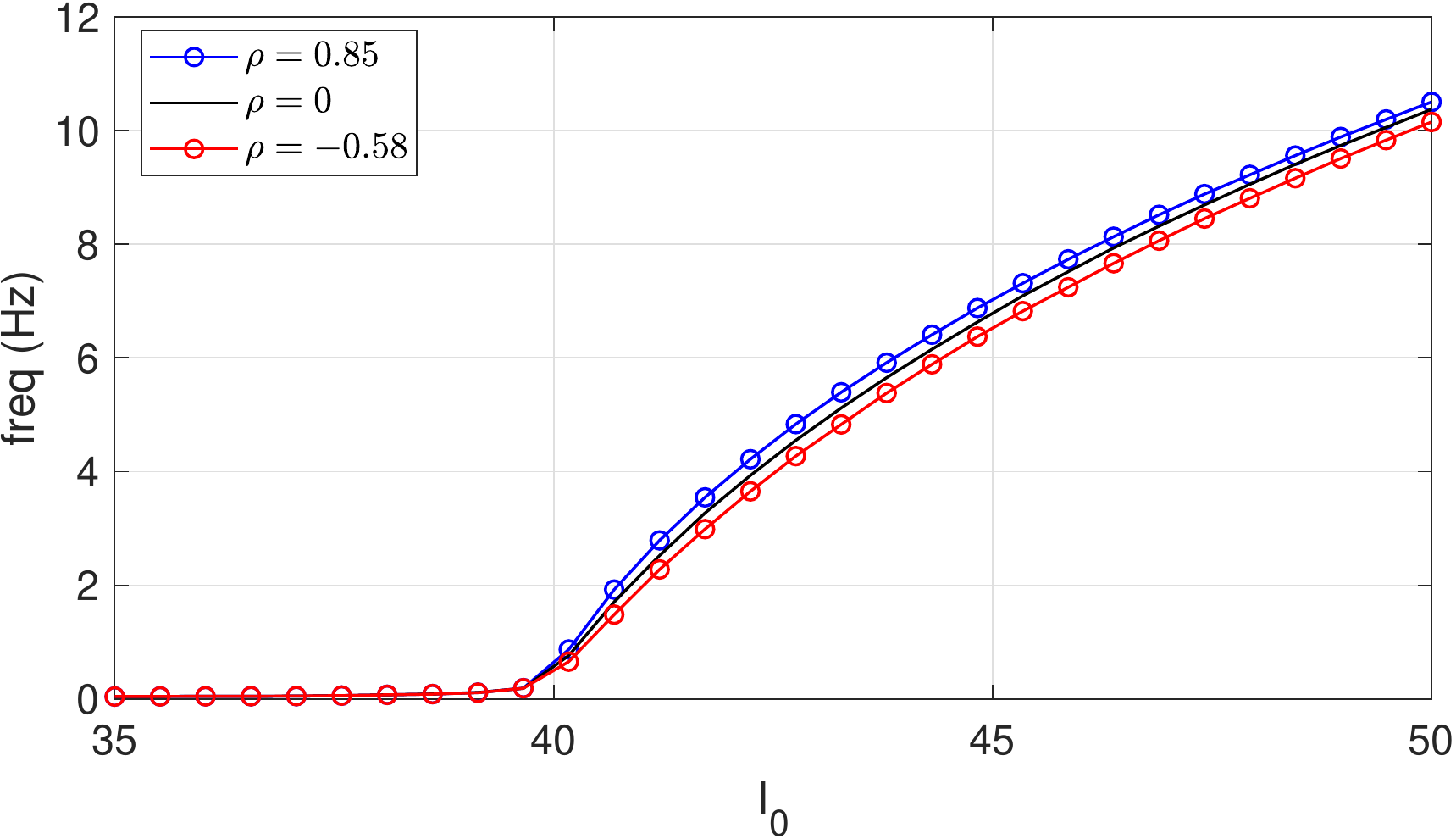}
\caption{Mean frequency versus $I_0$ for networks of Morris-Lecar neurons with 
inhibitory coupling.  $N=2000$.}
\label{fig:vIneg}
\end{figure}

For inhibitory coupling we replace $m_\infty(V_i)$ in~\eqref{eq:dsdt} by $w_\infty(V_i)$,
replace $V_{ex}-V_i$ in~\eqref{eq:dVdt} by $V_K-V_i$, and choose $\epsilon=10 mS/cm^2$.
Sweeping through $I_0$ for three different values of $\rho$ we obtain Fig.~\ref{fig:vIneg}, in qualitative
agreement with Fig.~\ref{fig:bistabD}.

\section{Motifs}
\label{sec:motif}
\begin{figure}
\includegraphics[scale=0.65]{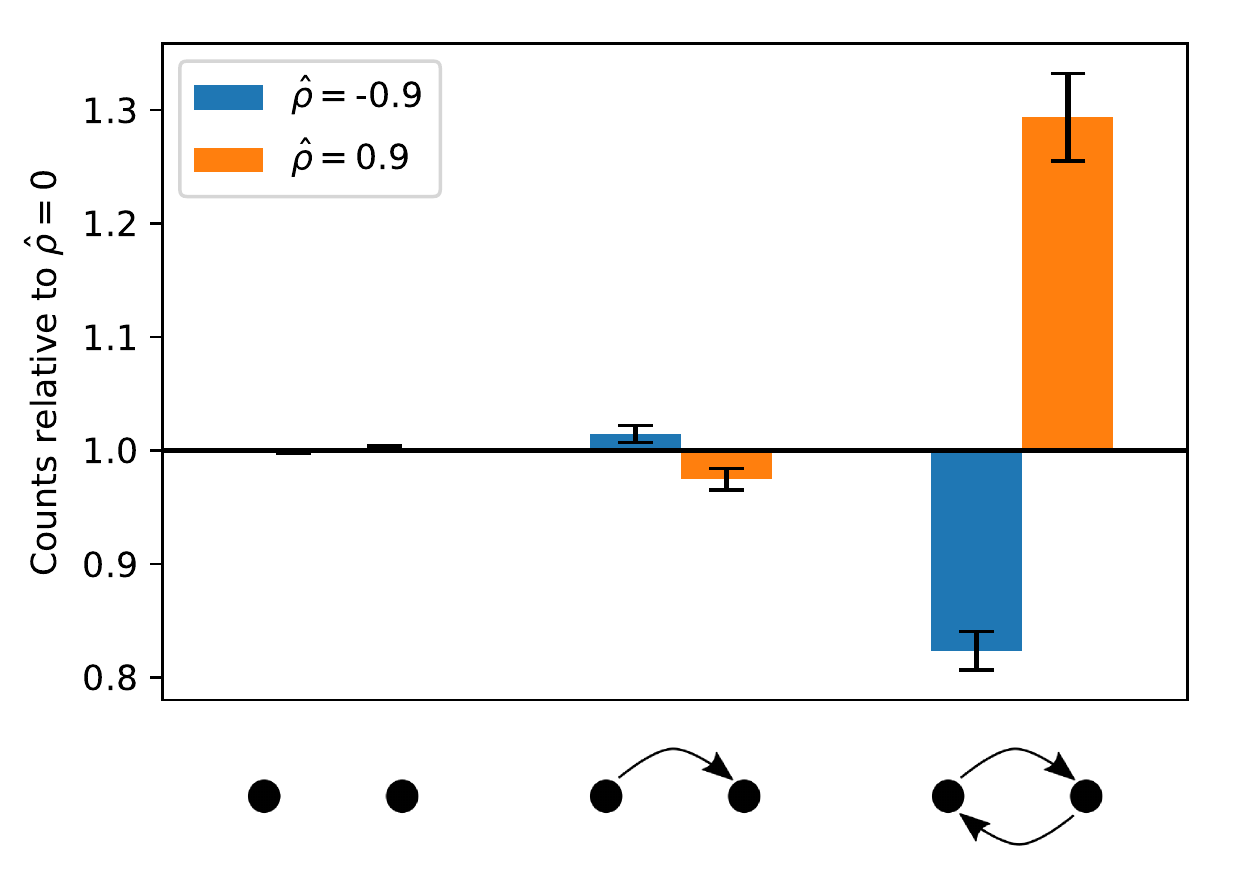}
\caption{Relative counts of order-2 motifs. We generate three networks at a time with $\hat{\rho}\in[-0.9, 0, 0.9]$ to compute motif frequencies and repeat this process 100 times.
Error bars indicate the standard deviation. Parameters are chosen as in Figure\ \ref{fig:Anet}.}
\label{fig:ord2}
\end{figure}

\begin{figure}
\includegraphics[scale=0.35]{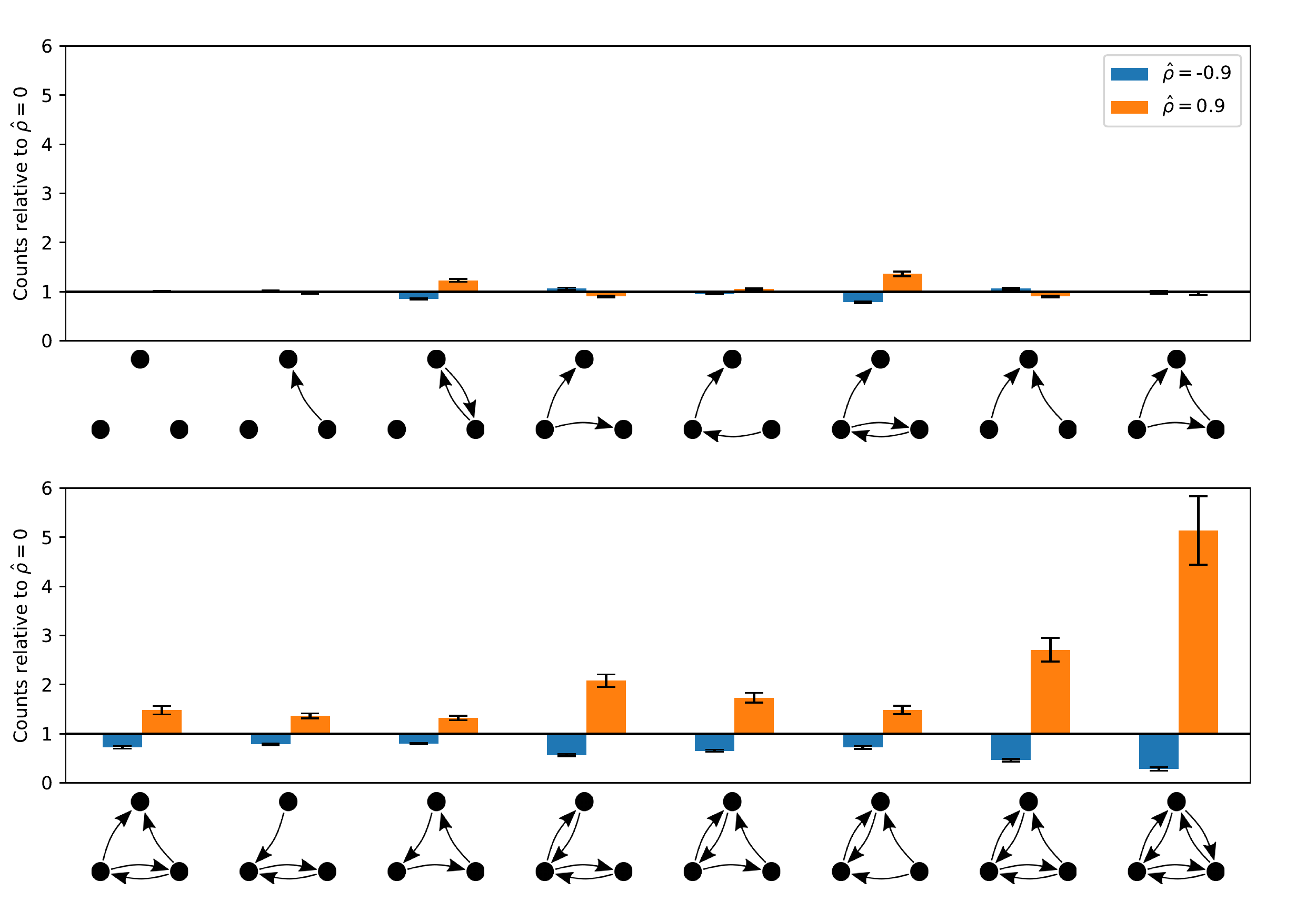}
\caption{Relative counts of order-3 motifs.}
\label{fig:ord3}
\end{figure}

A number of authors have found that ``motifs'' (small sets of neurons connected in a specific way)
do not occur in cortical networks in the proportions one would expect by 
chance~\cite{sonsjo05,Perber11}.
Some theoretical results relating the presence or absence of certain motifs to network
dynamics have been obtained~\cite{zhabev11,hutro13,ocklit15}. For networks whose
generation is described in Sec.~\ref{sec:ML} we counted the number of order-2 and
order-3 motifs (involving two or three neurons respectively), 
for negative, zero and positive values of $\rho$.
We compute the frequencies (amount) of order-2 motifs by counting the amount of 0's, 1's and 2's in the upper triangular part of $(A+A^T)$, where $A$ is the adjacency matrix and $T$ means transposed.
They refer to unconnected, unidirectional connected and reciprocal connected pairs of neurons, respectively.
For all 13 connected order-3 motifs we used the software ``acc-motif''\cite{acc-motif}.
The remaining three unconnected motifs have been counted by our own algorithm, i.e. looping through all neurons, we create for each a list of disconnected neurons and count among those order-2 motifs.
The results are shown in Figs.~\ref{fig:ord2} and~\ref{fig:ord3}, where counts are shown
relative to the numbers found for $\rho=0$.

In all motifs with at least one reciprocal connection between two neurons, we see that the number
of motifs goes up with positive $\rho$ and down with negative $\rho$. This can be understood
in an intuitive way: suppose $0<\rho$ and consider a neuron with a high out-degree. It is
likely to connect to a neuron with a high in-degree. But this second neuron will also have a high
out-degree and is therefore more likely to connect to the first neuron, which also has a high
in-degree, forming a reciprocal connection. Similarly, suppose $\rho<0$ and consider a neuron
with high out-degree. It is likely to connect to a neuron with high in-degree but low out-degree.
Thus it is unlikely that this second neuron will connect back to the first, which has a low
in-degree.

\section{Conclusion}
\label{sec:conc}
We have investigated the effects of correlating the in- and out-degrees of spiking neurons
in a structured network. We considered a large network of theta neurons, allowing us to exploit
the analytical results previously derived by~\cite{chahat17}, which give dynamics for
complex-valued order parameters, indexed by neurons with the same degrees. The states of interest
are steady states of these dynamics, and by using a Gaussian copula we were able to analytically
incorporate a parameter which controls the correlations between in- and out-degrees.
Numerical continuation was then used to determine the effects of varying parameters, 
particularly the degree correlation. In order to reduce the computational cost we introduced
the concept of ``virtual degrees'' allowing us to efficiently approximate sums with many terms
by sums with fewer terms.

For an excitatory network we found that increasing degree correlations had a similar effect
as increasing the overall strength of coupling between neurons, consistent with the
findings of~\cite{nykfri17,vegrox19}. Our results are also consistent with those
of~\cite{vashou12}, who found that negative correlations stabilised the low firing rate
state, as shown in Fig.~\ref{fig:bistabA}.
For inhibitory coupling we found that increasing degree
correlations slightly increased the mean firing rate of the network. Both of these
effects were reproduced in a more realistic networks of Morris-Lecar spiking neurons.

We also measured the relative frequency of occurence of order-2 and order-3 motifs as
within-degree correlations were varied and found that in all motifs with at least one reciprocal connection between two neurons, the number
of motifs is positively correlated with $\rho$. Several authors have linked motif
statistics to synchrony within a network~\cite{hutro13,zhabev11}, however a link between
motif statistics and firing rate, as observed here, seems yet to be developed.

We chose a Lorentzian distribution of the $\eta_i$ in~\eqref{eq:dtheta}, as many others
have done~\cite{ottant08}, in order to analytically evaluate an integral and derive~\eqref{eq:dbdt}.
However, we repeated the calculations shown in 
Figs.~\ref{fig:bistabA},~\ref{fig:bistabD},~\ref{fig:bistabB} and~\ref{fig:vIneg}
using a Gaussian distribution of the $\eta_i$ and found the same qualitative behaviour
(not shown). Regarding the parameter $n$ governing the sharpness of the function
$P_n(\theta)$, we repeated the calculations shown in Figs.~\ref{fig:bistabA}
and~\ref{fig:bistabD} for $n=5,\infty$ and obtained qualitatively the same results (not shown).
We used a Gaussian copula to correlate in- and out-degrees due to its analytical form,
but numerically investigated the scenarios shown in  Figs.~\ref{fig:bistabA}
and~\ref{fig:bistabD} for $t$ copulas and Archimedean Clayton, Frank and Gumbel copulas
and found the same qualitative behaviour (also not shown). 

For simplicity we used the same truncated power law distribution for both in- and 
out-degrees. However, the use of a Gaussian copula for inducing correlations
between degrees does not require them to be the same, so one could use the framework presented here
to investigate the effects of varying degree distributions~\cite{rox11}, correlated or not.

We also only considered either excitatory or inhibit\-ory networks, but it would be
straightforward to generalise the techniques used here to the case of both types
of neuron, with within-neuron degree correlations for either or both populations, though at the
expense of increasing the number of parameters to investigate.

\begin{acknowledgements}
This work is partially supported by the Marsden Fund Council 
from Government funding, managed by Royal Society Te Ap\={a}rangi. We thank Andrew Punnett
and Marti Anderson for useful conversations about copulas and Shawn Means for comments
on the manuscript. We also thank the referees for their helpful comments which improved the
paper.
\end{acknowledgements}

%
\section*{Conflict of interest}
The authors declare that they have no conflict of interest.




\end{document}